\documentstyle[12pt]{article}
\def\be{\begin{equation}}
\def\ee{\end{equation}}
\def\ba{\begin{eqnarray}}
\def\ea{\end{eqnarray}}
\def\ga{\mathrel{\raise.3ex\hbox{$>$\kern-.75em\lower1ex\hbox{$\sim$}}}}
\def\la{\mathrel{\raise.3ex\hbox{$<$\kern-.75em\lower1ex\hbox{$\sim$}}}}

\begin{document}

\begin{titlepage}
\pagestyle{empty}
\baselineskip=21pt
\rightline{UMN--TH--1805/99}
\rightline{TPI--MINN--99/31}
\rightline{hep-ph/9906331}
\rightline{June 1999}
\vskip.25in
\begin{center}

{\large{\bf Assisted Chaotic Inflation\\[1mm]
in Higher Dimensional Theories}}
\end{center}
\begin{center}
\vskip 0.5in

{Panagiota Kanti and Keith A. Olive}
\vskip 0.2in
{\it
{Theoretical Physics Institute, School of Physics and Astronomy, \\
University of Minnesota, Minneapolis, MN 55455, USA}}
\vskip 0.5in
{\bf Abstract}
\end{center}
\baselineskip=18pt \noindent
We address the problem of the large initial field values
in chaotic inflation and propose a remedy in the framework of the
so-called assisted inflation. We demonstrate that a 4-dimensional theory
of multiple, scalar fields with initial field
values considerably below the Planck scale, can give rise to inflation
even though none of the individual scalar fields are capable of driving
inflation. The problems arising from the presence of possible
non-renormalizable interactions is therefore removed.
 As a concrete example of a theory with multiple scalar
fields, we consider a $(4+d)$-dimensional field theory of a single,
non-interacting massive scalar field whose KK modes play the role of the
assisted sector. For the KK modes to assist inflation, the extra dimensions
must have a size larger than the inverse (4D) Planck
scale.  

\vspace*{25mm}
\begin{flushleft}
\begin{tabular}{l} \\ \hline
{\small Emails: yiota@physics.umn.edu, olive@mnhep.hep.umn.edu}
\end{tabular}
\end{flushleft}

\end{titlepage}
\baselineskip=18pt

\section{Introduction}
Barring an alternative, inflation is an essential component of the standard
big bang cosmological model.  Because there is no firm model of inflation,
particularly from a particle physics point of view, the literature is abound
with models of inflation \cite{rev}.  Without a doubt, the simplest of these
models is the chaotic inflationary scenario \cite{chaotic}.  This model is
based on a single scalar field and can be formulated either as a theory of a
massless scalar with a quartic self-coupling $\lambda$ or a non-interacting
massive scalar with mass $m$. In either variety, the magnitude of the
density perturbations produced by quantum fluctuations of the scalar field
during its evolution allows one to fix either the mass or quartic coupling,
$m \sim 10^{-5} M_P$ (where $M_P$ is the Planck mass) or $\lambda \sim
10^{-12}$. 

The initial conditions for chaotic inflation do not rely on an initial
thermal state.  Rather, chaotic inflation is based on
the simple assumption that the initial vacuum
energy density is of the Planck scale, that is $V(\phi) \simeq M_P^4$.
Inflation occurs for $\phi \ga M_P$ and its duration will be long enough
to solve the cosmological problems for $\phi >$ a few $M_P$. Of course,
the condition on the vacuum density implies $\phi \gg M_P$, and there
will in general be significantly more inflation than the necessary
minimum. It is precisely this condition on the initial values of the
inflaton, $\phi_0$, that can in principle be troublesome.
For example, in order to assume $\phi_0 \gg M_P$ and use only the
quadratic or quartic potential terms, one is implicitly assuming that
all non-renormalizable terms (whether at the tree level or arising at one
or more loops) are absent or severely suppressed\footnote{A more
complicated picture involving two or more scalar fields can relieve the
problem of extremely large initial field values \cite{lincon}.}\cite{em}.

Recently, it was shown \cite{liddle} that a system of several scalar
fields each with an exponential potential could drive power-law
inflation with a net power greater than the one achieved by a single
field. Indeed, it is possible to show that the inflation
driven by such a system of multiple fields is sufficient to solve the
cosmological problems, even if each of the individual fields
$\phi_i$ alone is not capable of doing so. This idea was extended to
quartic and quadratic potentials in \cite{yko1}, where it was shown that
the fine-tuning of the quartic coupling $\lambda \sim 10^{-12}$ could be
relaxed in the framework of a 4-dimensional theory of multiple 
scalar fields. However, it was also shown that these fields must not
be cross coupled, otherwise, their multiplicity actually
impedes inflation~\cite{yko1,maz}. The question of density perturbations
(both adiabatic and isocurvature) in assisted inflation was discussed
in \cite{mw}. 

In \cite{yko1}, we also argued that the source of a large number of
nearly identical scalar fields could be the compactification of a
(relatively) large extra dimension. In this case, although the
tower of KK states are heavily cross coupled, an attractor solution to the
equations of motion was found such that the fine-tuning of the quartic
coupling could be relaxed. Here, we will explore the effects of the large
number of scalar fields that arise from the compactification of a theory
containing a ($4+d$)-dimensional non-interacting massive
field\footnote{For recent work concerning inflation and large extra
dimensions, see \cite{large}. In \cite{ekoy}, KK-modes were used in
constructing inflationary models in the context of higher derivative
gravity models.}. In particular, we will focus on a remedy of another
potential problem concerning chaotic inflation, that of the large initial
conditions on the value of the inflaton field.  We will show that the
system of equations of motion will admit to  inflationary solutions for
which the initial field values of {\it all} the scalar fields is
significantly below the Planck scale, and as a result are not affected by
any possible non-renormalizable interactions.

\section{Assisted inflation and initial conditions}

Although we will be mainly interested in an $m^2 \phi^2$ chaotic
inflationary model, with the aim of relaxing the initial conditions on the
inflaton $\phi$, we begin by considering a general $m^2 \phi^2 +
\lambda \phi^4$ potential. In addition, let us consider a theory with
multiple, self-interacting scalar fields with a Lagrangian given by the
expression
\be
-{\cal L}= \sum_{i=1}^N \,\Biggl\{\frac 12\,\partial_\mu \phi_i\,
\partial^\mu \phi_i + \frac{1}{2}\,m^2\,\phi_i^2 + \frac{\lambda}{4!}\,
\phi_i^4\Biggr\}\,\,,
\label{orig}
\ee
with each field satisfying the equation of motion
\be
\ddot \phi_i + 3 H \dot \phi_i = -{dV_i \over d \phi_i}=-m^2\,\phi_i -
\frac{\lambda}{3!}\,\phi_i^3 \,\,.
\ee
The Hubble parameter is given by 
\be
H^2 = \frac{8\pi G}{3}\, \sum_{i=1}^N \,\Biggl\{ \frac 12\,\dot \phi_i^2 + 
\frac{1}{2}\,m^2\phi_i^2 + \frac{\lambda}{4!}\,\phi_i^4 \Biggr\}\,\,.
\ee

The equations of motion of any two scalar fields $\phi_i$ and
$\phi_j$ are solved by a simple scaling solution and the unique late-time
attractor of the system is given by $\phi_i=\phi_j$. In terms of this
solution, which is valid at all times, the Lagrangian can be written as
\cite{yko1}
\be
-{\cal L} = N\,\Biggl\{\frac 12\,\dot{\phi_1}^2 +
\frac{m^2}{2}\,\phi_1^2 + \frac{\lambda}{4!}\,\phi_1^4\Biggr\}
=\frac 12\,\dot{\tilde{\phi}}^2 + \frac{\tilde{m}^2}{2}\,\tilde{\phi}^2 +
\frac{\tilde{\lambda}}{4!}\,\tilde{\phi}^4\,\,,
\label{toy}
\ee
where
\be
\tilde{\phi}=\sqrt{N}\,\phi_1\,\,, \qquad \tilde{m}^2=m^2\,\,,
\qquad \tilde{\lambda}=\frac{\lambda}{N}\,\,.
\label{inflaton}
\ee
Thus, the above field redefinition allows us to rewrite the Lagrangian in
terms of a single scalar field with a canonical kinetic term.
Note that the mass of the renormalized scalar field, $\tilde{\phi}$, is
unchanged though its quartic coupling is reduced by a factor $N$. This
allows one to relax the fine-tuning of the quartic coupling and, at the
same time, remain consistent with the energy density fluctuations observed
in the micro-wave background.

As we noted in the introduction, 
if $\tilde{\phi}$ satisfies the initial condition, 
$\tilde{\phi}_0 \ga 3 (4.5) M_P$, the theory described by either the
quadratic (quartic) terms in (\ref{toy}) is  well known to lead to a
period of inflationary expansion of the universe with time-scale
\be
H \tau =\frac{2\pi\,\tilde{\phi}_0^2}{M_P^2} ~~
(\frac{\pi\,\tilde{\phi}_0^2}{M_P^2}) > 65\,\,.
\ee
However, the field $\tilde{\phi}$ that plays the role of the inflaton is only
an effective field constructed from the multiple, scalar fields $\phi_i$ of
the original theory (\ref{orig}). In terms of these fields, the initial
condition on $\tilde{\phi}$ translates to
\be
(\phi_i)_0 \ga \frac{3 (4.5) M_P}{\sqrt{N}}
\label{simres}
\ee
which means that, for a large number of multiple fields, the initial value
of each field $\phi_i$ can be considerably smaller than $M_P$, thus
relieving the concern regarding the role (or magnitude) of possible
non-renormalizable contributions to the Lagrangian. As a result, the
multiplicity of a theory with respect to the number of scalar fields that
it contains may resolve the problem of large initial conditions necessary
for chaotic inflation as long as the masses of these fields are
equal.  
%


%
\section{Kaluza-Klein reduction of a $\hat{m}^2\hat{\phi}^2$ theory}

As a possible source of a field theory with multiple scalar fields,
we consider a $(4+d)$-dimensional gravitational theory of the form 
\be
S_{4+d}= -\int d^{4+d}x\,\sqrt{G_{4+d}}\,\Biggl\{
\frac{M^{2+d}}{16\pi}\,R_{4+d}
+\frac{1}{2}\,G^{AB}_{4+d}\,\partial_A \hat{\phi}\, \partial_B \hat{\phi}
+\frac{1}{2}\,\hat{m}^2 \hat{\phi}^2 +\frac{\hat{\lambda}}{4!}\,
\frac{\hat{\phi}^4}{M^d}\Biggr\}
\ee
where $A, B=\{t, x_1, x_2, x_3, z_1, z_2,..., z_d\}$ and $M$ is the 
$(4+d)$-dimensional Planck mass. The Kaluza-Klein reduction of the above
theory will lead to the appearance of additional scalar fields,
the Kaluza-Klein modes, in the framework of the 4-dimensional effective
theory. If we assume that the extra $d$ dimensions are compactified over an
internal manifold with the size of every compact dimension parametrized by
$L$, we can Fourier expand the scalar field $\hat{\phi}$ along the compact
dimensions in the following way
\be
\hat{\phi}(x,z) = \hat{\phi}_0(x) + \hat{\phi}_z(x,z)=
\hat{\phi}_0(x) + \frac{1}{\sqrt{2}}
\sum_{\vec{n}} 
\Bigl( \hat{\phi}_{\vec{n}}(x) \,e^{i\frac{\pi\vec{n}}{L}\vec{z}} + 
\hat{\phi}_{\vec{n}}(x) \,e^{-i \frac{\pi\vec{n}}{L}\vec{z}} \Bigr) 
\ee
where $\vec{n}=\{n_1, n_2,..., n_d\}$ and $\hat{\phi}_0$ is the zero-mode
that depends only on non-compact coordinates. Note that there is a maximum
value of every $n_i$ which is defined by $n_i \leq N \simeq L M$,
corresponding to momenta $p_i = M$. In
\cite{yko1}, we assumed that the mass 
$\hat{m}$ was zero and that the potential of the 5-dimensional field
$\hat{\phi}$ was characterized by a quartic self-interaction term.
After compactification, the system
of the resulting Kaluza-Klein states is heavily cross coupled.
Nevertheless, we were able to solve for an attractor solution of the
system in terms of which the theory of multiple, cross coupled scalar
fields was mapped into a theory of a single, self-interacting, field
(through a quartic potential). The corresponding coupling constant was
given by the expression
\be
\lambda=\hat{\lambda}\,\biggl(\frac{M_5}{M_P}\biggr)^2
\ee
which relates the 4-dimensional, effective quartic coupling constant
and Planck mass with the corresponding quantities of the original
5-dimensional theory (the analysis of \cite{yko1} can be easily extended
to an arbitrary number of extra dimensions and the above formula remains
unchanged). Bearing in mind that the smallest possible value of the
5-dimensional Planck mass is $M_5 \sim 10^{-6} M_P$ for inflation
to occur~\cite{yko1}, it is easy to see that a suppression factor of
$10^{-12}$ comes directly from the ratio $(M_5/M_P)^2$ implying that
$\hat{\lambda}$ can be of ${\cal O}(1)$. Just as one can consider the
4-dimensional gravitational scale characterized by $M_P$, to be an
effective scale related to a fundamental, higher-dimensional energy
scale and the size of the relatively large, extra dimensions
by
\be
M_P^2=L^d\,M^{2+d} \,\,,
\label{planck}
\ee
the 4-dimensional quartic coupling constant $\lambda$ can be considered
as an effective coupling arising from the compactification of
a higher-dimensional, fundamental theory. 

A simple set of (non-cross-coupled) scalar fields with quartic
self-couplings, as described in the previous section, allows one to
scale-down of the initial field values while still providing chaotic
inflation so long as eq. (\ref{simres}) is satisfied. The fact that the
system of the multiple Kaluza-Klein states in \cite{yko1} were
non-trivially coupled, precludes such a solution. Instead of a solution
of the form given in eq. (\ref{inflaton}) defining the inflaton, we found
that for large $N$, $\tilde \phi \simeq \sqrt{1+ {1\over N}}\,\phi_0$,
where $\phi_0$ is the zero KK mode of the theory~\footnote{There is a
solution for which $\tilde{\phi} \simeq \sqrt{1+N}\,\phi_0$, but
in this case the fine-tuning problem of the quartic coupling is not
resolved and $\lambda=\hat{\lambda}$.}. Here, we limit our
discussion to a non-interacting
massive scalar field in (4+d) dimensions and set
$\hat \lambda = 0$. The resulting 4-dimensional, effective action is then
\ba
S_{eff} &=& -\int d^{4}x\,\sqrt{g}\,\Biggl\{\frac{M_P}{16\pi}\,R
+\frac{1}{2}\,\partial_\mu \phi_0 \, \partial^\mu \phi_0 + 
 \frac{1}{2}\,\sum_{\vec{n}=1}^N
\partial_\mu\phi_{\vec{n}}\, \partial^\mu \phi_{\vec{n}}+
\nonumber\\[3mm]
&~& \hspace*{2.5cm}\frac{1}{2}\,m^2 \phi_0^2 +
\frac{1}{2}\,\sum_{\vec{n}=1}^N (m^2+m_{\vec{n}}^2)\,\phi_{\vec{n}}^2
\Biggr\}\,\,, \label{final}
\ea
where we have made use of the following relations
\be
\phi=L^{d/2}\,\hat{\phi}\,\,, \qquad m^2=\hat{m}^2
\ee
between the $(4+d)$-dimensional and the 4-dimensional quantities, of the
fundamental relation (\ref{planck}) and of the definition 
\be
m_{\vec{n}}^2=\frac{\vec{n}^2 \pi^2}{L^2}
\ee
for the Kaluza-Klein masses. In addition, the compactification of the extra
dimensions  gives rise to a dilaton-like field $\gamma$ and a gauge field
$A_\mu$  defined as $G_{ii}=e^\gamma$, for $i=5,...,d$, and $G_{\mu
d}=e^\gamma A_\mu$, respectively. We will, however, ignore the
gauge field and assume that the dilaton-like field remains fixed~\cite{ac}.
Note that, unlike the $\hat{\lambda} \hat{\phi}^4$ case~\cite{yko1}, the
resulting 4-dimensional system we consider here, is still a simple set of
non-interacting, massive scalar fields. We will show that this system
leads to the relaxation of the large initial conditions according to the
recipe of section 2. 

The zero-mode field $\phi_0$ and the higher Kaluza-Klein excitations
$\phi_{\vec{n}}$ satisfy the equations of motion
\ba
\ddot{\phi}_0 + 3H \dot{\phi}_0 &=& -m^2 \phi_0\,\,, \label{zero}\\[3mm]
\ddot{\phi}_{\vec{n}} + 3H \dot{\phi}_{\vec{n}}
&=& -(m^2 + m_{\vec{n}}^2)\,\phi_{\vec{n}}\,\,,\label{high}
\ea
respectively. Note that, although the bare mass $m^2$ has remained unchanged
after the compactification of the extra dimensions, the appearance of a
Kaluza-Klein mass for the higher modes of the theory leads, in principle,
to the existence of a finite tower of scalar fields with {\em different}
masses. 

In principle, the multiple scalar fields of the 4-dimensional effective
theory (\ref{final}) can have arbitrary initial values. 
If we assume that all of the KK fields have initial values larger
than $3 M_P$,  each field of the theory can drive inflation 
independently of the presence of all other fields.
As an illuminating example,
we consider the case of two scalar fields, $\phi_1$ and $\phi_2$, with
masses $m_1$ and $m_2 \gg m_1$, respectively, that satisfy the following
equations of motion 
\ba
\ddot{\phi}_1 + 3H \dot{\phi}_1 &=& -m_1^2\,\phi_1\,\,, \\[3mm]
\ddot{\phi}_2 + 3H \dot{\phi}_2 &=& -m_2^2\,\phi_2\,\,.
\ea
Since both of the scalar fields have initial values larger than $3M_P$,
the slow roll-over conditions are satisfied for each of them and the
energy density of the system is dominated by their mass terms. If we ignore
the second derivatives of the fields with respect to time and rearrange
their equations of motion, the following scaling solution is found
\be
\frac{\dot{\phi}_1}{\dot{\phi}_2}=\frac{m_1^2}{m_2^2}\,
\frac{\phi_1}{\phi_2}\,\Rightarrow\,
\phi_1(t) \sim \phi_2(t)^{\,m_1^2/m_2^2}\,\,.
\label{scaling}
\ee
From the above relation, and since $m_1^2/m_2^2 \ll {\cal O}(1)$, we may
easily conclude that the light field $\phi_1$ evolves much slower towards
its minimum value than the heavy field $\phi_2$. 

In the present case, where a tower of Kaluza-Klein states with masses
ranging from $m^2$ to $m^2+M^2$ satisfy the slow roll-over conditions, 
the Universe may undergo one or more periods of inflation depending
on the initial field values, $\phi_i(0)$. Because of the scaling
(\ref{scaling}), the final period of inflation is typically driven by the
lightest of the scalar fields under consideration (see e.g.
\cite{graziani}). For certain initial field configurations, however,
quantum fluctuations may drive heavier scalar fields back to
configurations which will subsequently inflate \cite{fluc}.  We will
return to this issue below.

We are most interested in the case when all of the scalar fields
have initial values less than
$3 M_P$. In this case, the slow roll-over conditions can not be satisfied
for any of the individual fields and, as a  result, none of them can drive
inflation alone. The equations of motion of the
Kaluza-Klein fields (\ref{zero})-(\ref{high}) preclude a scaling
solution that holds between different fields, at all times, as long as
the masses of the fields are different. Moreover, the unique late-time
attractor of the system is the trivial one, $\phi_i=\phi_j=0$. Thus, each
field evolves towards its minimum independently of all the others, whose
presence become manifest only through their contribution to the Hubble
parameter. The duration of the roll-over phase for each field is
proportional to its initial value and inversely proportional to its mass. 

However, there is a subset of Kaluza-Klein fields which can be considered
to have approximately the same mass and whose evolution, although not
different from the picture described above, may give rise to an inflationary
period for the universe. For the lightest of the Kaluza-Klein fields,
the mass term $m^2_{\vec{n}}$ generated by the compactification of the
extra dimensions can be considered negligible compared to the bare mass
$m^2$. As a result, these fields can be grouped together, as in the model
described in section 2, and form an effective inflaton field,
$\tilde{\phi}$. For a large number of fields participating in the
construction of
$\tilde{\phi}$, the initial value of the inflaton may exceed the
threshold of $3 M_P$ producing inflation while the values of the
constituent fields remain well below the aforementioned threshold. 

If we generalize the chaotic initial condition, $V(\phi) \sim M_P^4$, to
the higher dimensional theory,  we may write that
$\hat{V}(\hat{\phi}) \sim M^{4+d}$. Then, in terms of 4-dimensional
quantities, we obtain
\be
V(\tilde{\phi})=L^d \hat{V}(\hat{\phi}) \sim M_P^2 M^2 \sim 
m^2 \tilde{\phi}^2\,\,.
\label{ref}
\ee
If $\tilde{\phi}$ is to drive inflation, we must assume that $\tilde{\phi}
\geq M_P$ and $m \sim 10^{-5} M_P$. Then, eq. (\ref{ref}) provides the
constraint $M \geq 10^{-5} M_P$ on the higher dimensional Planck mass.
Furthermore, the value of
$M$ determines the number of Kaluza-Klein fields that can be considered to 
have approximately the same mass. Comparing the terms $m^2$ and
$m^2_{\vec{n}}$, it is easy to conclude that the number of states that can
be considered to have the same mass $m$ must satisfy the constraint
\be
\vec{n}^2_m < \frac{m^2}{M^2}\,\Biggl(\frac{M_P}{M}\Biggr)^{4/d} 
\simeq \frac{m^2}{M^2}\,(N_d)^{2/d}\,\,,
\label{condition}
\ee
where $N_d \equiv (N+1)^d \simeq N^d = M_P^2/M^2$ is the total number of
Kaluza-Klein states in the theory. For $d=1$,  the actual
number $N_m$ of Kaluza-Klein states that have approximately the same
mass and, thus, participate in the construction of the inflaton field, is
given by $N_m = \sqrt{\vec{n}_m^2}$. For the smallest possible
value of the higher dimensional Planck mass, $M \sim 10^{-5} M_P$,
$N_m \simeq N_d \sim 10^{10}$ while for $M \sim 10^{-3} M_P$,
$N_m \simeq 10^{-2} N_d \sim 10^{4}$. In the former case, all of the KK
fields participate in assisting inflation and the result is impressive.
From eq.~(\ref{inflaton}), the initial values of the constituent fields
$\phi_{\vec{n}}$ can be as much as $10^5$ times smaller than
$M_P$ (i.e. of order the 5-dimensional Planck scale) and still be able to
produce inflation. When the value of $M$ is increased to
$10^{-3} M_P$, only $1\%$ of the KK fields participate. However,
even in the latter case, the relaxation of the large initial conditions is
still significant since the initial values of the constituent fields can,
now, be approximately 100 times smaller than $M_P$. 

For more than one extra
dimensions, the right-hand-side
of eq.~(\ref{condition}) becomes considerably smaller when compared to the
case $d=1$. However, there is a degeneracy of states
for $d>1$ that results in the existence, by construction, of
multiple states with the same mass and, to a large
extent, compensates the reduction in the rhs of  eq.~(\ref{condition}).
The number of states in general will be $N_m \propto
({\vec{n}^2_m})^{d/2}$. Therefore the number of states is
$N_m \propto (m/M)^d N_d$, and for $M \sim m$, the number of states is the
same in any dimension.  As a result, the construction of the inflaton field
and the relaxation of its large initial conditions is still feasible, for
an arbitrary number of extra dimensions, although less efficient as
$d$ increases if $M > m$.

Let us note, that, during the inflationary period driven by the inflaton
light field  $\tilde{\phi}$, scalar field quantum fluctuations~\cite{fluc}
are generated which may affect the evolution of the heavy KK fields which
would otherwise evolve towards their minimum without producing inflation. 
These quantum fluctuations may increase the initial values of the heavy
fields above the threshold of $3 M_P$ enabling them to drive inflation.
If we assume that $m_i^2 \ll H^2$, the quantum fluctuations grow as
$H^3 t/4\pi^2$ to a limiting value of
\be
\phi_i^2 \simeq \frac{2}{3}\,\frac{m^4\tilde{\phi}^4_0}{M_P^4 (m^2 +
m_i^2)}\,\,,
\ee
where $i$ denotes all the heavy Kaluza-Klein fields with masses equal to
or larger than $m^2$. If the slow roll-over conditions are satisfied for
any of the heavy fields,  they could drive a new period of inflation. This
is potentially problematic since they would produce an unacceptably large
size for the density perturbations. As long as we restrict the initial
value of the inflaton by
\be
\phi_i^2 \la 9 M_P^2 \qquad {\rm or} \qquad 
\tilde{\phi}_0 \la 3^{\,3/4} M_P (M_P/m)^{1/2} \simeq 720 M_P\,\,,
\ee
late periods of inflation driven by the heavy fields would not occur.
These conditions are easily satisfied if we assume that all of the light
fields have initial values suitably below the Planck scale.
\paragraph{}


\section{Conclusions}

Although the chaotic inflationary scenario is undoubtly the
simplest model for inflation, it relies on the assumption that
the initial values of the inflaton are large (compared to
$M_P$) in order to obtain a period of rapid
expansion long enough to solve existing
cosmological problems. Moreover, the fundamental parameters of
such a theory, being either the mass $m$ of a non-interacting
scalar field or the quartic self-coupling $\lambda$ of the
inflaton, must be extremely small in order to ensure
that the density perturbations produced during inflation have
the right order of magnitude. In a previous
article~\cite{yko1}, we demonstrated that a 4-dimensional theory
of multiple scalar fields can lead to chaotic inflation where
the coupling constant $\lambda$ is small due to the
multiplicity of the theory with respect to the scalar fields. 
As a concrete example of such a theory, we considered a
5-dimensional field theory where the Kaluza-Klein modes of the
5-dimensional scalar field played the role of the multiple
fields. However, no remedy for the problem of large initial
conditions was found in the context of the above theory due to
the cross-coupling terms between the KK modes.

Here, we focused on the problem of large initial conditions. 
We demonstrated that a 4-dimensional field theory of
multiple, self-interacting scalar fields can both resolve 
the fine-tuning problem of the quartic coupling
and relax the initial conditions on the value of the
inflaton field. The method is based on the construction of an
effective field -- the inflaton -- from the original, multiple
fields of the theory \cite{liddle}. In this case, the inflaton has the
same mass as the constituent fields but with an enhanced field value and
a considerably weaker self-interaction. The method becomes more
effective as the number of fields that participate in the
construction of the inflaton becomes larger. As a result, for a
large number of fields present in the theory, the quartic coupling
constant assumes naturally a small value in agreement to the COBE
data. Moreover, even for initial values of the constituent fields
many orders of magnitude smaller than $M_P$, the value of the
inflaton field can exceed the threshold of few $M_P$ and produce
inflation.

As a concrete example of a theory with multiple scalar fields, we
considered a $(4+d)$-dimensional field theory of a single, massive,
non-interacting scalar field. Upon performing a Kaluza-Klein
compactification of the extra dimensions, a 4-dimensional, effective
theory of multiple, massive, non-interacting fields emerged. The
appearance of a Kaluza-Klein mass for each one of the higher modes
of the theory led to the existence of a finite tower of scalar fields
with different masses. However, depending on the value of the 
higher-dimensional Planck mass $M$, many of these fields can
be considered to have approximately the same mass and, thus, give rise 
to an effective field that plays the role of the inflaton. For 
$M \sim 10^{-5} M_P$ or $M \sim 10^{-3} M_P$, the initial values of
the original fields of the theory can be $10^{-5}M_P$ or
$10^{-2}M_P$ respectively,  and still
enable the inflaton field to drive inflation. This removes any
potential problem concerning the presence of non-renormalizable
interactions in the theory.

{\large\bf Acknowledgments}
 This work was supported in part by 
DOE grant DE-FG02-94ER40823 at Minnesota.

\end{document}